\documentstyle[twocolumn,aps,psfig]{revtex}

\begin{document}

\twocolumn[\hsize\textwidth\columnwidth\hsize\csname @twocolumnfalse\endcsname
% Title Page
%

\title{Ferromagnetism in the Hubbard model on fcc--type lattices}

\author{Martin Ulmke}
\address{Theoretische Physik III, 
Universit\"at Augsburg, D--86135 Augsburg, Germany.
ulmke@physik.uni-augsburg.de}

\date{\today}
\maketitle

\begin{abstract}
The Hubbard model on fcc--type lattices is studied in the dynamical
mean--field theory of infinite spatial dimensions. At intermediate
interaction strength finite temperature Quantum Monte Carlo calculations 
yield a second order phase transition to a highly polarized, metallic
ferromagnetic state. 
The Curie temperatures are calculated as a function of electronic density
and interaction strength.
A necessary condition for ferromagnetism is a density
of state with large spectral weight near one of the band edges.


\end{abstract}

\vskip2pc]

More than 30 years ago, the Hubbard model was introduced to 
describe band magnetism in transition metals, in particular the  
ferromagnets Fe, Co and Ni \cite{Gutzwiller63,Hubbard63,Kanamori63}.
However, for a long time it appeared to be a generic 
model rather for \em anti--\em ferromagnetism than for ferromagnetism. 
At half filling (one electron per site) on bipartite lattices
antiferromagnetism emerges in both weak and strong coupling perturbational
approaches and, in particular, antiferromagnetism is tract\-able
by renormalization group methods \cite{Shankar94}.
By contrast, ferromagnetism is a non--trivial
strong  coupling phenomenon which cannot be investigated 
by any stand\-ard perturbation theory.
In fact, our knowledge on the possibility of ferromagnetism in the
Hubbard model is still very limited.
Only in a few special cases was the Hubbard model proven to have
ferromagnetic order. The first rigorous result showing a fully polarized
ground state, the theorem by Nagaoka \cite{Nagaoka66},
is only valid for a single hole added to the half filled band in the limit
of infinite on--site repulsion. 
In the limit of low electronic density a saturated
ferromagnetic ground state has been found recently
in the one dimensional Hubbard model 
on zigzag chains \cite{Mueller-Hartmann95,Tasaki95}.
Further, the existence of a ground state with net polarization has been
proven for the half filled band case on bipartite lattices with asymmetry
in the number of sites per sublattice \cite{Lieb89}, and in 'flat--band' 
systems \cite{Mielke93}.

Due to the development of new numerical algorithms and powerful computers
the problem of ferromagnetism in the Hubbard model has recently become
accessible to numerical investigations of finite systems, at least in reduced
dimensions $(d=1,2)$. 
In the presence of a next nearest neighbor hopping, 
polarized ground states were found very recently
in $d=1$ \cite{Daul97} and 
on square lattices in the case of a van Hove singularity
at the Fermi energy \cite{Hlubina97}.

The rigorous and numerical results mentioned above show that the  
stability of ferromagnetism is intimately linked with the structure of the 
underlying lattice and the kinetic energy (i.e.~the hopping) of the electrons.
This fact has also been observed by exact variational bounds for the stability
of saturated ferromagnetism 
\cite{Shastry90,Fazekas90,Mueller-Hartmann91etc,Hanisch95},
and within approximative methods \cite{Herrmann97etc}.
Generally, lattices with closed loops 
and with frustration of the competing anti\-ferro\-magnetism 
(non--bipartite lattices) are expected to support ferromagnetism.
Non--bipartite lattices have an asymmetric density of states (DOS) and thus
a peak away from the center of the band. 
Indeed, it has already been observed by the inventors of the model  
\cite{Gutzwiller63,Hubbard63,Kanamori63} that 
a DOS with a peak near one of the band edges is favorable for ferromagnetism.
Hence, the fcc lattice is expected to be a good candidate
for ferromagnetism because i) it has a highly asymmetric DOS
and ii) antiferromagnetism is frustrated.

Since numerical studies of finite systems would allow only very small fcc 
lattices, we study the Hubbard model in the dynamical mean--field theory 
(DMFT) which becomes exact 
in the limit of infinite dimensions, $d=\infty$ \cite{Metzner89etc,Georges96}.
We will show that an appropriate, non--singular kinetic energy is able
to induce metallic ferromagnetism in the single band Hubbard model for 
intermediate values of the interaction.
 
In the limit $d\to\infty$, 
the system is reduced to a dynamical single site problem 
\cite{Janis91} equivalent
to an Anderson impurity model complemented by a self--consistency condition
\cite{Georges92,Jarrell92}. 
Still, it cannot be solved analytically, and to avoid further 
approximations we employ a finite temperature Quantum Monte Carlo method 
\cite{Hirsch86}. 
Unlike traditional mean--field theories, the DMFT takes quantum fluctuations 
fully into account. Spatial fluctuations are neglected -- an approximation
which becomes exact 
in the limit of large coordination number, $Z$. For the three dimensional 
fcc lattice we have $Z=12$. The DMFT has proven to be 
a powerful and reliable tool for the study of three dimensional 
correlated Fermi systems both for thermodynamical as well as dynamical 
properties \cite{Georges96,Pruschke95}.
Spectral properties are obtained by analytic continuation of the imaginary
time data using the Maximum Entropy method \cite{Jarrell96}.
The solution of the self--consistent single site problem provides the
local self energy $\Sigma_{\sigma n}\equiv \Sigma_\sigma(i\omega_n)$
($\omega_n:$ Matsubara frequencies) from which the local Green function is
obtained by a Dyson equation:
\begin{equation}
G_{\sigma n} = \frac{1}{L} \sum_{\bf k} 
\frac{1}{z_{\sigma n}-\epsilon({\bf k})}
 = \int dE \frac{N_0(E)}{z_{\sigma n}-E}
\label{Gl1}
\end{equation}
with $z_{\sigma n}=i\omega_n+\mu-\Sigma_{\sigma n}$. Here,
$\epsilon({\bf k})$ are the single particle energies, 
$N_0(E)$ is the non--interacting DOS, and $L$ is the number of lattice sites.
Investigations of the Hubbard model on the hypercubic lattice in $d=\infty$
did not show indications for ferromagnetism for finite $U$ 
\cite{Georges96,Pruschke95},
however in a recent study at $U=\infty$
a ferromagnetic instability was found within the non--crossing approximation
\cite{Obermeier97}. 

A non--trivial generalization of the fcc lattice in $d$ dimensions 
\cite{Mueller-Hartmann91etc} (see also \cite{Uhrig96})
is the set of all points with integer cubic coordinates summing up to an even
integer. It is a non--bipartite Bravais lattice for any dimension $d>2$. 
Each point has $Z=2d(d-1)$ nearest neighbors defined by all vectors $\bf x$ 
which can be written in the form $\bf x = \pm \bf{e}_i \pm \bf{e}_j$, 
with two different cubic unit vectors
$\bf{e}_i$ and $\bf{e}_j$ ($i,j=1,\cdots,d$).
The energy dispersion reads
\begin{equation}
 \epsilon({\bf k}) = 
4 t \sum_{1\leq i < j\leq d} \cos{k_i} \cos{k_j} 
+ 2 t' \sum_{i=1}^d \cos{2 k_i}
\label{Gl2}
\end{equation}
where we allow a next nearest neighbor hopping $t'$.
In the particular case $t'=t/2$ we find 
$\epsilon({\bf k})=t \; \epsilon^2_{hc}({\bf k})-3t$
where $\epsilon_{hc}({\bf k})$ is the energy dispersion of the 
hypercubic lattice.
This relation leads to an inverse square--root divergency of the DOS
at the lower band edge in any dimension.
In the limit $d=\infty$ this divergency occurs for any $t'$, and
with the proper scaling $t=1/\sqrt{Z}$ \cite{Metzner89etc} the 
DOS can explicitly be calculated as \cite{Mueller-Hartmann91etc}:
\begin{equation}
N_0^{d=\infty}(E) = e^{-(1+\sqrt{2} E)/2}/\sqrt{\pi (1+\sqrt{2}E)} \; .
\label{Gl3}
\end{equation}
$N_0^{d=\infty}$ was also used by Uhrig's \cite{Uhrig96} calculation of the 
single spin--flip energy of the fully polarized state. There,
the ferromagnet turned out to be stable against a single spin--flip 
over a wide density regime.

In $d=3$ for $0<t'<t/2$ 
the divergency is absent, and the DOS has a typical $\sqrt{E-E_b}-$behavior
at the lower edge $E_b=\epsilon(\pi,\pi/2,0)=-4t+2t'$.
For $t'=0$ there is a weak, logarithmic divergency at $E_b$.
The main effect of $t'$ on the DOS is to induce a broad peak with 
much spectral weight near (but not right at) the lower edge. 
In (\ref{Gl3}) and in the following the energy scale is set by the variance
of the non--interacting DOS.
In $d=3$ for $0<t'<t/2$ the total band width is $W=16t+4t'$. 
By fixing the variance of the DOS, $v=\sqrt{12t^2+6t'^2}\equiv 1$, the 
band width ranges from 4.618 for $t'=0$ to 4.899 for $t'=t/2$.
Comparing with real 3d--transition metals, e.g.~Ni, one can roughly 
identify our energy scale with 1eV.

We will consider the following two cases: i) strictly $d=\infty$, 
i.e.~using the DOS (\ref{Gl3}) and ii) $d=3$ lattices, i.e.~using the
DOS of the the three dimensional lattice in (\ref{Gl1}).
In practice we perform finite sums over the $\bf{k}$--vectors corresponding
to a finite three dimensional lattice. To keep the finite size error 
at least one order of magnitude smaller than the statistical errors,
the number of lattice sites (respectively $\bf{k}$--vectors) 
has to be of the order
of $10^5$ to $10^7$ depending on temperature. The computer time for this
summation is still much smaller than for the Monte Carlo sampling.   

To detect a ferromagnetic instability we determine 
the temperature dependence of the uniform static 
susceptibility, $\chi_F$, from the two particle correlation functions
\cite{Ulmke95a}. In addition, we calculate commensurate magnetic 
susceptibilities and the compressibility (charge susceptibilty).
No charge instability is observed in the parameter regime under consideration.

At an intermediate interaction strength of $U=4$ we find the 
ferromagnetic response to be strongest around quarter filling ($n\simeq 0.5$).
$\chi_F$ obeys a Curie--Weiss law (see Fig.~1 for case i)) and the Curie 
temperature $T_c$ can safely be extrapolated from the zero of $\chi_F^{-1}$
to a value of $T_c=0.051(2)$ at $n=0.58$ \cite{Note1}.
Below $T_c$ the magnetization $M$ grows rapidly, reaching more than
80\% of the fully polarized value ($M_{max}=n=0.58$) at the lowest
temperature which is only 30\% below $T_c$.
The three data points $M(T)$ (Fig.~1) are consistent with a 
Brillouin function with the same critical temperature 
of $T_c=0.05$ and an extrapolated full polarization at $T=0$.
A saturated ground state magnetization is also consistent with 
the single spin--flip energy
of the fully polarized state \cite{Uhrig96}.
\begin{figure}[h]
\vspace*{-30mm}
\hspace*{-30mm}
\psfig{file=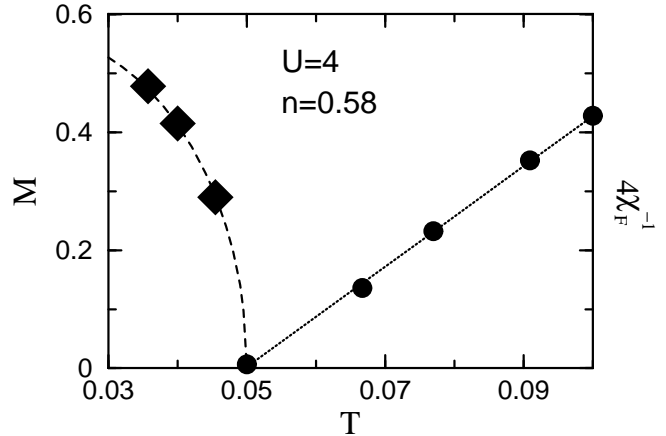,width=135mm,angle=-90}
\vspace{-18mm}
\caption{Magnetization, $M$ (diamonds), and inverse uniform static
susceptibility, $\chi_F^{-1}$ (circles; values multiplied by a factor of 4
to use the same scale) for $U=4$ and $n=0.58$. Error--bars are of the size of
the symbols or smaller. Dotted line is a linear fit to $\chi_F^{-1}$,
dashed line a fit with a Brillouin function to $M$. (Note that the circle
at $T=0.05$ is a data point, not an extrapolation.)}
\end{figure}

The Curie temperatures are obtained from the Curie--Weiss plots for
different values of interaction $U$ and electronic densities $n=N_{el}/L$, 
leading to the phase diagram Fig.~2 for the case i). 
At low temperatures the system is ferromagnetic over a wide density regime.
$T_c$ increases with $U$ and the upper critical density $n_c$ seems to 
approach $n=1$. 
The values of $n_c$ as a function of $U$ 
are consistent with the single spin--flip results \cite{Uhrig96}. 
Note that the Stoner criterion, $UN(\mu)>1$, would give $n_c>1$ for $U>1.5$. 
For low densities the Curie temperature becomes very small and seems to 
vanish close to $n=0$. 
Antiferromagnetism is not expected on the fcc lattice in high dimensions
because the difference of the numbers of not frustrated bonds and frustrated 
bonds is only of the order of $d$ resulting in an effective field of the 
order of $t^2 d\propto 1/d$ \cite{Mueller-Hartmann95etc}.
\begin{figure}[h]
\vspace*{-32mm}
\hspace*{-30mm}
\psfig{file=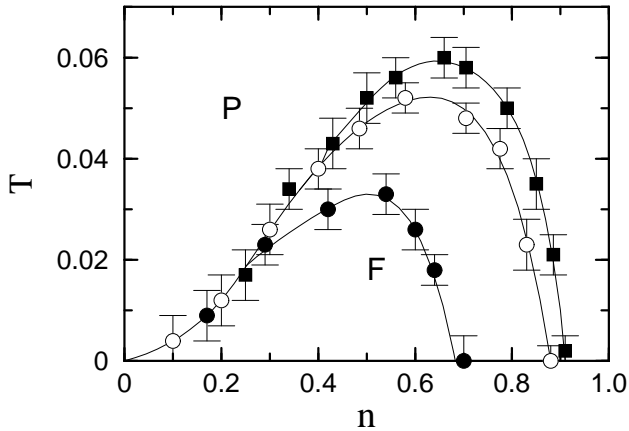,width=135mm,angle=-90}
\vspace{-18mm}
\caption{$T_c$ vs.~$n$ phase diagram of the Hubbard model on the 
$d=\infty$ fcc lattice with $U=2$ (full circles), 
$U=4$ (open circles), and $U=5$ (squares). Lines are guides to the eye.
}

\end{figure}
\begin{figure}[h]
\vspace*{-34mm}
\hspace*{-30mm}
\psfig{file=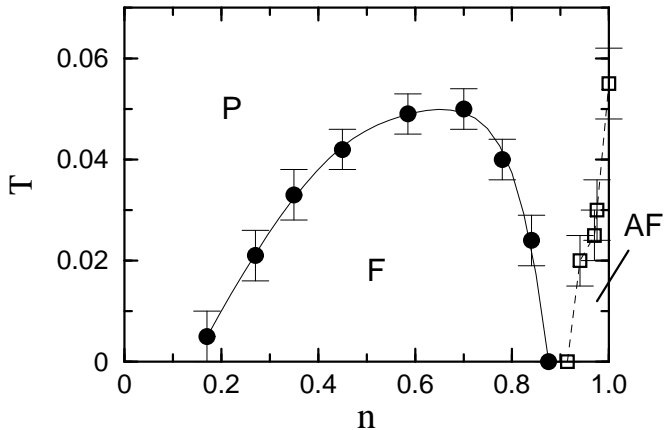,width=135mm,angle=-90}
\vspace{-18mm}
\caption{Phase diagram of the Hubbard model on the 
$d=3$ fcc lattice with $t'=t/4$ and $U=6$.
Circles: ferromagnetic, squares antiferromagnetic phase boundary.}
\end{figure}

In case ii) we first studied the pure fcc lattice, i.e.~$t'=0$. However
in this situation, and for intermediate $U-$val\-ues 
($U\leq 6)$ the Curie--Weiss extrapolation would lead to very small 
Curie temperatures, numerically indistinguishable from $T_c=0$. 
If we, however, introduce a next nearest neighbor hopping term of $t'=t/4$ the
situation is again similar to the case i). The phase diagram for $U=6$ 
(Fig.~3)  
shows a large ferromagnetic regime with an optimal density about $n=0.65$. 
For $U=4$ no ferromagnetism was found for temperatures
accessible to the Monte Carlo technique.
In contrast to the case i) now there seems to exist a finite lower critical
density. This might be due to the fact that the DOS has no divergency at
the band edge but a foot. 
An additional feature is the antiferromagnetic 
phase very close to half filling. 
We found an instability at ${\bf k}=(\pi,0,0)$, corresponding
to ferromagnetic planes with alternating orientation in direction 
perpendicular to the planes. 
In addition there seem to be instabilities 
at ${\bf k}=(\pi/2,\pi/2,\pi/2)$ and ${\bf k}=(\pi/2,\pi/2,0)$,
however only at $n=1$ and very low temperatures $(T<0.01)$. 
At $n=1$ the N\'{e}el temperature is (numerically) degenerate with an
instability at ${\bf k}=(\pi,\pi/2,0)$. Below half filling the latter is 
more strongly suppressed.

The single particle spectra for spins parallel
and antiparallel to the net magnetization (Fig.~4 for case i)) are
apparently metallic since both spectra have a finite value
at the Fermi level ($\omega=\mu$). This also holds in the paramagnetic phase
(not shown).
The minority spin spectrum (dotted line in Fig.~4) is not only shifted 
to higher frequencies but has also a little foot at low energies.
This foot contains about 50\% of the spectral weight below $\mu$ 
and looses weight with increasing polarization.
The minority spin spectrum shows a pronounced upper band around
$\omega-\mu\approx U=4$ with a developing (pseudo) gap.
The majority spin spectrum, however, is only slightly affected by the
interaction. Here, the weight of the upper band is very small since 
the Pauli principle makes it unlikely for the majority spin--electrons 
to hop over occupied sites.
\begin{figure}[h]
\vspace*{-30mm}
\hspace*{-30mm}
\psfig{file=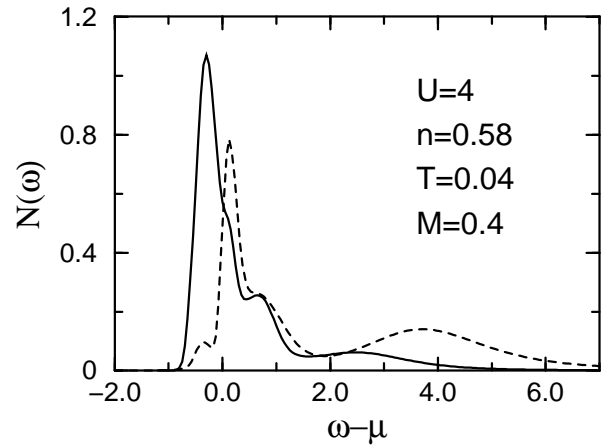,width=135mm,angle=-90}
\vspace{-18mm}
\caption{Single particle spectrum, $N(\omega)$, for $U=4$, $n=0.58$, 
and $T=0.04$. Full (dotted) line: majority (minority) spin spectrum.} 

\end{figure}
The exchange splitting, defined by the difference of the peak positions, 
is $\Delta\approx 0.43$ which is significantly
smaller than the shift $\Delta^{Stoner}=UM\approx 2$ in Stoner theory. 
Further, Stoner theory would give a Curie temperature 
$T_c^{Stoner}\approx 0.5$ which is one order of magnitude too high.
Thus, the mean--field Stoner theory cannot describe the different energy
scales correctly.
Identifying our energy scale with 1eV, the resulting Curie temperature
corresponds to $T_c\approx 600K$. We note that this value and the large 
ratio $\Delta/T_c>8$ -- which will even somewhat increase for $T\to 0$ --
are in good agreement with experimental values on fcc Ni 
($T_c=627K$, $\Delta/T_c\approx 10$) \cite{Wohlfarth80}.
The agreement is quite surprising considering the neglect of
band degeneracy and inter--orbital interactions
favoring ferromagnetism (Hund's coupling).

While the saturated ferromagnet has zero interaction energy
($U\langle d \rangle=0$), the mobility of the electrons and hence the
kinetic energy is very poor, since electrons cannot hop over occupied sites. 
The paramagnetic state, on the other hand, also tries to minimize 
double occupancies, $\langle d \rangle$, and thus has a small 
kinetic energy, too.
For example at $U=4$ and $n=0.58$, we find $U \langle d \rangle<0.01$ 
for $T<0.06$. Hence the gain in interaction energy 
cannot account for the relatively high Curie temperatures.
This raises the question: under which circumstances has the fully
polarized ferromagnet a lower \em kinetic \em energy than the paramagnet?
In a one--particle picture at $T=0$ the kinetic energy difference is:
\begin{equation}
{\Delta E} = \int^{E_{\mbox{\scriptsize\em Ferro}}} dE \, E \, N_0(E)
-  2 \int^{E_P} dE \, E \, \tilde N(E) 
\label{deltae}
\end{equation}
where the Fermi levels $E_{\mbox{\scriptsize\em Ferro}}$ and $E_P$ 
have to determined by
the density $n$.
In (\ref{deltae}), $\tilde N(E)$ is the (unknown) DOS of the correlated 
paramagnet. 
To get an idea of the energy difference we consider the following simple
picture: since double occupancies are effectively suppressed by $U$,
the band will consist of two parts separated by an energy $U$. 
For a given $\uparrow$--electron the average number of nearest neighbors
which are \em not \em occupied by $\downarrow$--electrons is 
$q=1-n_{\downarrow}=1-n/2$ in the paramagnet, and hence the effective width
of the lower band will be reduced by that very factor $q$.
The simplest assumption for the lower part of the spectrum is
$\tilde N(E)= N_0(E/q)$
i.e.~a renormalization of the energies \cite{Hubbard-I}.
For the DOS (\ref{Gl3}), 
${\Delta E}(n)$ is negative for all densities $0<n<1$ and vanishes for
$n=0$ and $n=1$.
The density dependence of $|{\Delta E}(n)|$ resembles strongly the
$T_c(n)$ data (Fig.~2). In particular $|{\Delta E}(n)|$ becomes maximal
at $n=0.735$ with $|{\Delta E}|_{max}=0.0747$. 
For a DOS with the same shape but a divergency at the \em upper \em edge, 
${\Delta E}$ is always positive, i.e.~the ferromagnet is totally unstable.
Apparently this simple treatment of the correlated paramagnet 
describes qualitatively the stability of the ferromagnet depending
on the structure of the non--interacting DOS, and gives even the correct
energy scale for $T_c$.
To check this approximation one can also estimate $\Delta E$ by 
replacing the second integral of (\ref{deltae}) by a numerical integration
over the finite temperature DOS in the paramagnetic 
solution. For $n=0.58$, $U=4$, and $T=0.04$ (cf.~Fig.~4, but now in 
the paramagnetic solution) 
we obtain $\Delta E'\approx -0.04(1)$ \cite{Note2} which is in 
reasonable agreement with the simple approximation above
($\Delta E =  -0.064$).

In summary, we presented a numerical proof of metallic ferromagnetism
in the single band Hubbard model in the dynamical mean--field theory.
A density of states with large spectral weight near one of the band
edges is an essential ingredient for ferromagnetism.
This condition goes far beyond the Stoner criterion, $UN_0(\mu) >1$, because
it is the band renormalization of the \em interacting \em paramagnet 
-- a manifest many body effect -- 
which determines the stability of ferromagnetism.

The author acknowledges helpful discussions and correspondence with 
K.~Held, E.~M\"uller-Hartmann, R.~Scal\-ettar, J.~Schlipf, G.~Uhrig, 
D.~Vollhardt, 
and J.~Wahle. 
He thanks A.~Sandvik for kindly providing his Maximum Entropy program.
This work was supported in part by a grant from
the ONR, N00014--93--1--0495 and by the DFG.


\begin{references}
%

\bibitem{Gutzwiller63}
M.~C.~Gutzwiller, Phys.~Rev.~Lett.~{\bf 10}, 159 (1963).

\bibitem{Hubbard63}
J.~Hubbard, Prog.~Roy.~Soc.~London A {\bf 276}, 238 (1963).

\bibitem{Kanamori63}
J.~Kanamori, Prog.~Theor.~Phys.~{\bf 30}, 275 (1963).

\bibitem{Shankar94}
R.~Shankar, Rev.~Mod.~Phys.~{\bf 66}, 129 (1994).

\bibitem{Nagaoka66}
Y.~Nagaoka, Phys.~Rev.~{\bf 147}, 392 (1966).

\bibitem{Mueller-Hartmann95} E.~M{\"u}ller-Hartmann, 
J.~Low.~Temp.~Phys.~{\bf 99}, 349 (1995).

\bibitem{Tasaki95}
For similar lattices with two minima in the dispersion 
ferromagnetism was found in the case of a half filled subband.
H.~Tasaki, Phys.~Rev.~Lett.~{\bf 75}, 4678 (1995).

\bibitem{Lieb89} E.~H.~Lieb, Phys.~Rev.~Lett.~{\bf 62}, 1201 (1989).

\bibitem{Mielke93}
A.~Mielke and H.~Tasaki, Commun. Math. Phys. {\bf 158}, 341 (1993).

\bibitem{Daul97} S.~Daul and R.~M.~Noack, Z. Phys.~B {\bf 103}, 293 (1997). 
\bibitem{Hlubina97}
R.~Hlubina, S.~Sorella, and F.~Guinea, 
Phys.~Rev. Lett.~{\bf 78}, 1343 (1997).

\bibitem{Shastry90}
B.~S. Shastry, H.~R. Krishnamurthy, and P.~W. Anderson,
Phys.~Rev.~B {\bf 41}, 2375 (1990).

\bibitem{Fazekas90}
P.~Fazekas, B.~Menge, and E.~M{\"u}ller-Hartmann,
Z. Phys.~B {\bf 78}, 69 (1990).

\bibitem{Mueller-Hartmann91etc}
E. M{\"u}ller-Hartmann, in {\em Proc.~V Symp.~Phys.~of Metals}, edited by E.
  Talik and J. Szade (Poland, 1991), p. 22.

\bibitem{Hanisch95} 
T.~Hanisch, B.~Kleine, A.~Ritzl, and E.~M{\"u}ller-Hartmann,
Ann.~Physik {\bf 4}, 303 (1995). T.~Hanisch, G.~Uhrig,
and E.~M{\"u}ller-Hartmann, Phys.~Rev.~B {\bf 56}, 13960 (1997).

\bibitem{Herrmann97etc}
T.~Herrmann and W.~Nolting, Solid State Commun.~{\bf 103}, 351 (1997);
M.~Potthoff, T.~Herrmann, and W.~Nolting, preprint.

\bibitem{Metzner89etc}
W. Metzner and D. Vollhardt, Phys.~Rev.~Lett.~{\bf 62}, 324 (1989); 
D. Vollhardt, in {\em Correlated Electron Systems}, 
edited by V.~J.~Emery (World Scientific, Singapore, 1993), p.~57.

\bibitem{Georges96}
A.~Georges, G.~Kotliar, W.~Krauth, and M.~Rozenberg,
Rev.~Mod.~Phys.~{\bf 68}, 13 (1996).

\bibitem{Janis91}
V.~Jani\v{s}, Z.~Phys.~B {\bf 83}, 227 (1991).

\bibitem{Georges92}
A.~Georges and G.~Kotliar,
Phys.~Rev.~B {\bf 45}, 6479 (1992).

\bibitem{Jarrell92}
M.~Jarrell, Phys.~Rev.~Lett.~{\bf 69}, 168 (1992).

\bibitem{Hirsch86}
J.~E. Hirsch and R.~M. Fye,
Phys.~Rev.~Lett.~{\bf 56}, 2521 (1986).

\bibitem{Pruschke95}
T.~Pruschke, M.~Jarrell, and J.~K.~Freericks,
Adv.~Phys. {\bf 44}, 187 (1995).

\bibitem{Jarrell96}  For a review see M.~Jarrell and J.~E.~Gubernatis, 
Phys.~Rep.~{\bf 269}, 133 (1996).

\bibitem{Obermeier97} T.~Obermeier, T.~Pruschke, and J.~Keller, 
Phys.~Rev.~B {\bf 56}, R8479 (1997).

\bibitem{Uhrig96}
G.~S. Uhrig,
\newblock Phys.~Rev.~Lett.~{\bf 77}, 3629 (1996).

\bibitem{Ulmke95a}
M.~Ulmke, V.~Jani\v{s}, and D.~Vollhardt,
\newblock Phys.~Rev.~B {\bf 51}, 10411 (1995).

\bibitem{Note1} The data in Fig.~1 are obtained for one small but fixed
discretization $\Delta\tau=0.25$ in imaginary time. The error in $T_c$ is 
predominantly due to the additional extrapolation to $\Delta\tau\to 0$.

\bibitem{Mueller-Hartmann95etc}
E.~M{\"u}ller-Hartmann (private communication); Burkhard Kleine, Dissertation,
  University of Cologne (1995).

\bibitem{Wohlfarth80}
E.~P.~Wohlfarth, in {\em Ferromagnetic Materials, Vol.~1}, edited by
E.~P.~Wohlfarth (North Holland, Amsterdam, 1980), p.~1, and references
therein.

\bibitem{Hubbard-I} 
In the limit of large $U$ this turns out to be equivalent to the Hubbard-I
approximation \cite{Hubbard63}.

\bibitem{Note2} The error
in $\Delta E'$ is due to uncertainties in the 
analytic continuation and in the electronic density.
The temperature broadening of the the Fermi--Dirac function
is negligable at $T=0.04$.


\end{references}
\end{document}